\documentclass{aa}
\usepackage{graphicx}
\begin{document}
   \title{Photometry of dissolving star cluster candidates}

   \subtitle{The cases of NGC~7036 and NGC~7772\thanks{Based on observations
carried out at Mt Ekar, Asiago, Italy},\thanks{Photometry is only available
in electronic form at the CDS via anonymous ftp to {\tt cdsarc.u-strasbg.fr (130.79.128.5} or via {\tt http://cdsweb.u-strasbg.fr/cgi-bin/qcat?J/A+A//}}
}

   \author{Giovanni Carraro}

   \offprints{G. Carraro}

   \institute{Dipartimento di Astronomia, Universit\`a di Padova,
              Vicolo Osservatorio 2, I-35122 Padova, Italy\\
              \email{giovanni.carraro@unipd.it}
             }

   \date{Received November 2001; accepted}

   \abstract{ We present CCD UBVI observations obtained in the field
of the two previously unstudied dissolving open cluster candidates NGC~7036
and NGC~7772. Our analysis suggests that  both the objects are Open
Cluster Remnants (OCR).\\
NGC~7036 is an open
cluster remnant with a core radius of about 3-4 arcmin. 
We derive for the first time 
estimates of its fundamental parameters. We identify 17 likely members 
which define a group  of stars at 1 kpc from the Sun, 
with a low reddening E$(B-V) \approx 0.1$,
and with an age of about 3-4 Gyr.\\
As for NGC~7772, we identify 14 likely members, which define
a group of stars with a very low reddening (E$(B-V) \approx 0.03$), 1.5 Gyr old
and located about 1.5 kpc from the Sun.
   }

   \maketitle
%

\section{Introduction}
The dynamical evolution and the final fate of open star clusters in the Milky 
Way is nowadays a very active research field.
Open star clusters are weakly bound objects with a typical 
lifetime of less than a Gyr (Dutra \& Bica 2000, Bergond et al. 2001), which ultimately
depends on the initial mass of the cluster, the birthplace and
the fraction of primordial binaries (de la Fuente Marcos 1998, 2001).\\
Recently, Bica et al. (2001) draw the attention on a sample of
high Galactic latitude ($b > 15^{o}.$)  
star clusters presumably in a advanced stage of dynamical evolution, which
they baptized {\it Probable Open Cluster Remants (POCR)}.
The prototype of this class of objects is NGC~6994 (M~73), recently studied
by Bassino  et al. (2001) and Carraro (2001), who performed the first
multicolor photometric studies of this cluster, but
arrived at opposite conclusions on the nature of this object. 
While Bassino et al. suggest that M~73 is the remnant of a star cluster, Carraro
(2000) proposes that it is just a chance alignment
of four bright stars.\\
Although it is a difficult task to unravel the nature of a star concentration
basing only on photometry, it however remains the first necessary
step. Indeed sometimes the Color Magnitude Diagrams (CMDs) and Color-Color
Diagrams (CCDs) are sufficient to disentangle between a real bound system or a
random enhancement of stars (Carraro \& Patat 1995, Piatti \& Clari\`a 2001).
In many cases however, photometry cannot help to decide unambiguously about
the nature of a star concentration: in this situation radial velocities
and/or proper motions studies are necessary (Baumgardt 1998, Baumgardt et al.
2000, Odenkirchen \& Soubiran 2002).\\
The census provided by Bica  et al. (2001) lists 20 candidate dissolving
open clusters, some of which completely unstudied.
This is the case of NGC~7036 and NGC~7772, two high latitude objects
traditionally  considered genuine open clusters, which are the subject of the
present study.\\
The basic idea of this paper is to present the first photometric study
of these clusters and to provide a list of probable members stars to
be further studied with high resolution spectroscopy.\\
A nice example of that is the recent spectroscopic follow-up
of NGC~6994 by Odenkirchen \& Soubiran (2002), who confirmed
Carraro (2000) suggestions that this object is a chance
alignment of four bright stars.\\

In Sect.~2 we briefly present the observations and data reduction.
Sects.~3 and 4 illustrate our results for NGC~7036 and NGC~7772, and,
finally, Sect.~5 draws some conclusions and suggests further lines
of research.

\begin{table}
\caption{Basic parameters of the observed objects.
Coordinates are for J2000.0 equinox}
\begin{tabular}{ccccc}
\hline
\hline
\multicolumn{1}{c}{Name} &
\multicolumn{1}{c}{$\alpha$}  &
\multicolumn{1}{c}{$\delta$}  &
\multicolumn{1}{c}{$l$} &
\multicolumn{1}{c}{$b$} \\
\hline
& $hh:mm:ss$ & $^{o}$~:~$^{\prime}$~:~$^{\prime\prime}$ & $^{o}$ & $^{o}$ \\
\hline
NGC~7036       & 21:10:02 & +15:31:06 & 64.55  & -21.44\\
NGC~7772       & 23:51:46 & +16:14:48 & 102.74 & -44.27\\
\hline\hline
\end{tabular}
\end{table}

\section{Observations and Data Reduction}

Observations were carried out with the AFOSC camera at the 
1.82~m Copernico telescope of Cima Ekar (Asiago, Italy), in the photometric
night of October 21 , 
2001. AFOSC samples a $8^\prime.14\times8^\prime.14$ field in a  
$1K\times 1K$ thinned CCD. The typical seeing was between 2.0  
and 2.5 arcsec.  
 
The basic data of the studied clusters are summarized in Table~1, whereas
the details of the observations are listed in Table~2.
The covered regions are shown in Fig~1 and~6, where two DSS\footnote
{Digital Sky Survey, {\tt http://archive.eso.org/dss/dss}} maps
are presented for NGC~7036 and NGC~7772, respectively.
 
The data has been reduced by using the IRAF\footnote{IRAF  
is distributed by the National Optical Astronomy Observatories, 
which are operated by the Association of Universities for Research 
in Astronomy, Inc., under cooperative agreement with the National 
Science Foundation.} packages CCDRED, DAOPHOT, and PHOTCAL. 
The calibration equations obtained by observing Landolt(1992) 
PG~02331 field 3 times along the night, are: 
	\begin{eqnarray}  
\nonumber 
u \! &=& \! U + 3.707\pm0.054 + (0.201\pm0.060)(U\!-\!B) + 0.58\,X \\  
\nonumber 
b \! &=& \! B + 1.545\pm0.006 - (0.068\pm0.007)(B\!-\!V) + 0.29\,X \\  
\nonumber 
v \! &=& \! V + 0.893\pm0.029 + (0.010\pm0.011)(B\!-\!V) + 0.16\,X \\  
\nonumber 
i \! &=& \! I + 1.624\pm0.017 - (0.051\pm0.015)(V\!-\!I) + 0.08\,X \\ 
	\label{eq_calib} 
	\end{eqnarray} 
where $UBVI$ are standard magnitudes, $ubvi$ are the instrumental  
ones, and $X$ is the airmass. The standard stars in this field
provide a very good color coverage, being $-0.329 \leq (B-V) \leq 1.448$.
For the extinction coefficients, 
we assumed the typical values for the Asiago Observatory
(Desidera et al. 2001).
Photometric global errors have been estimated following Patat
\& Carraro (2001). For the $V$ filter,  
they amount at 0.03, 0.05 and 0.07 at $V\approx$
12.0, 16.0 and 20.0, respectively.

\begin{table} 
\tabcolsep 0.30truecm 
\caption{Journal of observations of NGC~7036 and NGC~7772 
and standard star field PG~02331 (October 21, 2001).} 
\begin{tabular}{cccc} 
\hline 
\multicolumn{1}{c}{Field}    & 
\multicolumn{1}{c}{Filter}    & 
\multicolumn{1}{c}{Time integration}& 
\multicolumn{1}{c}{Seeing}         \\ 
      &        & (sec)     & ($\prime\prime$)\\ 
  
\hline 
 NGC~7036   &     &              &      \\ 
            & $U$ &  240,600     &  2.2 \\ 
            & $B$ &  30,300      &  2.3 \\ 
            & $V$ &  30,300      &  2.3 \\ 
            & $I$ &  30,200      &  2.3 \\ 
 NGC~7772   &     &              &      \\ 
            & $U$ &  300         &  2.2 \\ 
            & $B$ &  30,120      &  2.4 \\ 
            & $V$ &  5,15,60     &  2.3 \\ 
            & $I$ &  5,15,60     &  2.3 \\
 PG~02331   &     &              &      \\ 
            & $U$ &  60,120,60   &  2.0 \\ 
            & $B$ &  60,60,120   &  2.1 \\ 
            & $V$ &  15,15,30,30 &  2.0 \\ 
            & $I$ &  30,30,15    &  2.1 \\
\hline 
\end{tabular} 
\end{table}

   \begin{figure}
   \centering
   \resizebox{\hsize}{!}{\includegraphics{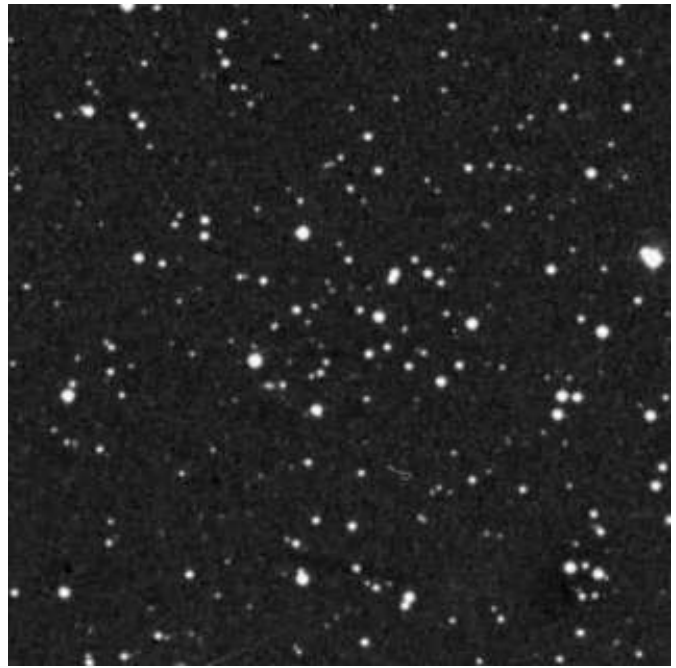}} 
   \caption{ A DSS red map of the covered region in the field
of NGC~7036. North is up, East on the left.}
    \end{figure}

\section{NGC~7036}
This object has never studied before, and it is not
included in the Lyng\aa~ (1987) open cluster catalog.
NGC~7036 (see Fig~1) appears as a weak star enhancement, confined within 
a 3 squared arcmin area in a relatively rich stellar field. 

   \begin{figure}
   \centering
   \resizebox{\hsize}{!}{\includegraphics{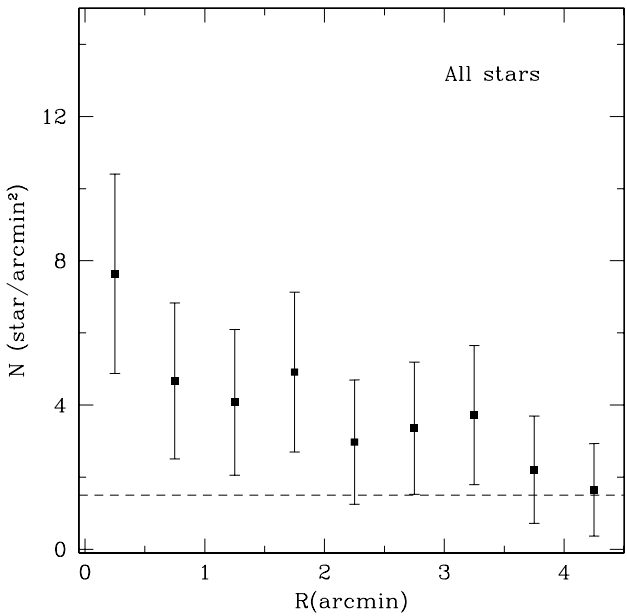}} 
   \caption{ Star counts in the field of 
of NGC~7036 as a function of the radius. The dashed line is the
field number density estimate provided by Bica et al. (2001).}
    \end{figure}

\subsection{Star counts}
The first signature of the possible presence
of a star clusters is recorded in star counts.
Bica et al. (2001) compared star counts in the region of NGC~7036
with a Galactic model and DSS maps, showing that NGC~7036 
significantly emerges from the surrounding field.\\
This is confirmed by stars counts based on the present study.\\
We derived the surface stellar density by performing star counts
in concentric rings around star \#17 (selected as 
approximate cluster center)
and then dividing by their
respective surfaces. The final density profile and the corresponding
poissonian error bars are depicted in Fig.~2.
In this figure we take into account all the measured stars.\\
The surface density decreases smoothly  over all the region 
we covered, suggesting that the cluster has a radius about 3.5-4 arcmin,
and actually emerges significantly from the background.
We count 5 stars brighter than $V~=13.7$ in a field of about 9 squared arcmin,
exactly the same number reported by Bica  et al. (2001, Table~1). However
we notice that these stars are not in the cluster central region, but
populate a sort of ring between 2 and 3 arcmin. The cluster center is
on the other hand populated by fainter stars presumably belonging
to the field. 
This can be interpreted as an indication of the dissolution the cluster
is undergoing.

   \begin{figure}
   \centering
   \resizebox{\hsize}{!}{\includegraphics{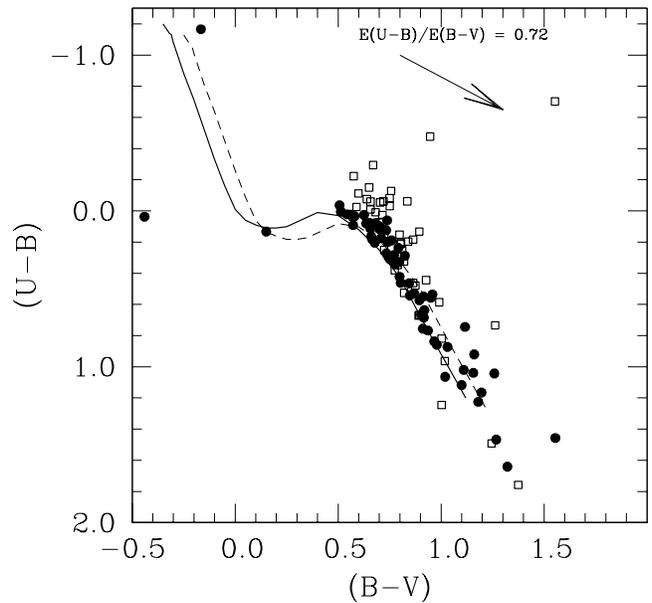}} 
   \caption{ Color-color diagram for all the stars in the field
of NGC~7036 having $UBV$ photometry. Filled circles indicate 
stars brighter than $V$ = 17, whereas open squares 
indicate the remaining stars.
The solid line is a Schmidt-Kaler (1982) empirical ZAMS, whereas the dashed
line is the same ZAMS, but shifted by E$(B-V)$~=~0.1.}
    \end{figure}

   \begin{figure}
   \centering
   \resizebox{\hsize}{!}{\includegraphics{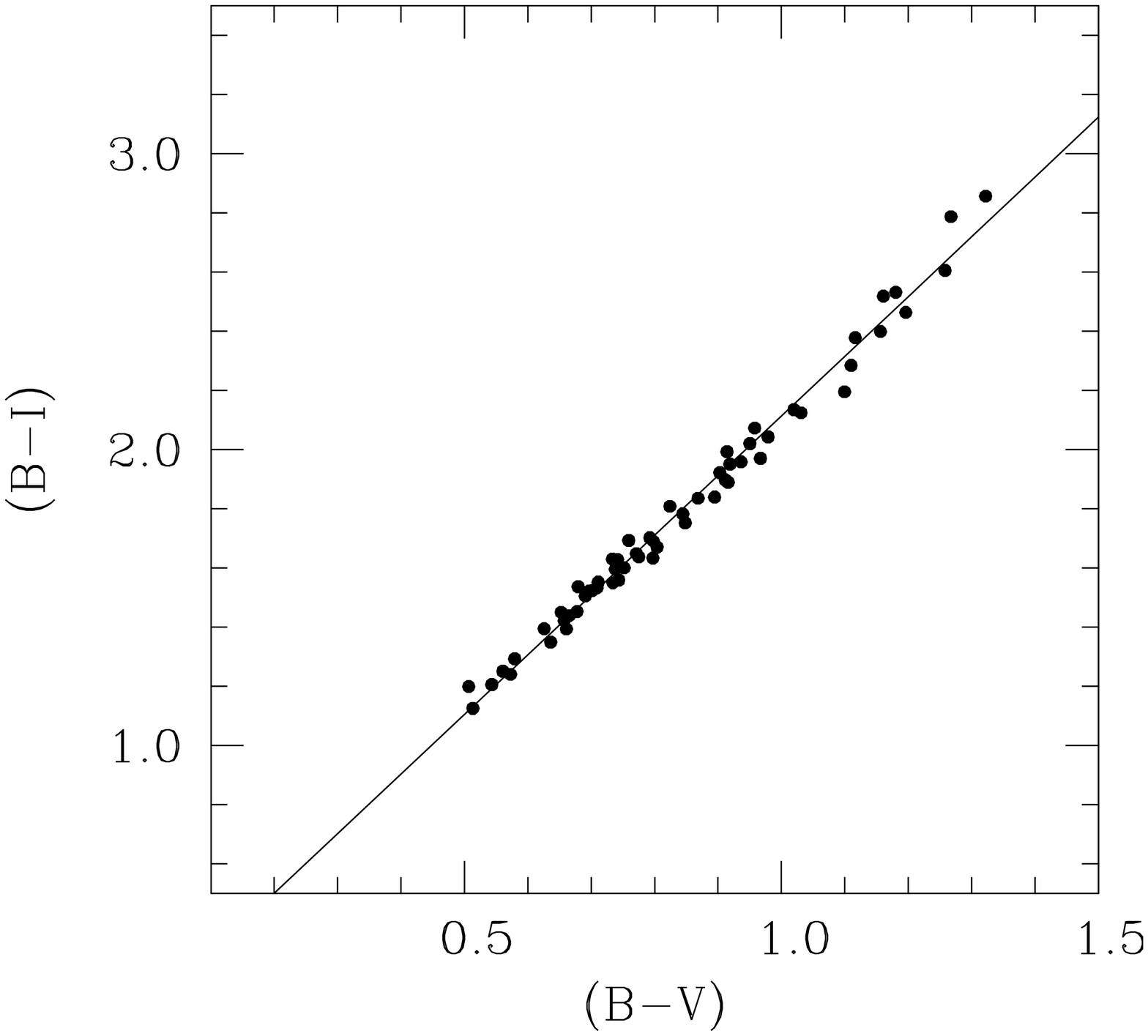}} 
   \caption{ NGC~7036 stars brighter than $V$~=~17 in the $(B-V)$ vs
$(B-I)$ plane.}
    \end{figure}

\subsection{Color-Color and Color-Magnitude Diagrams}
In order to better understand the nature of NGC~7036, we constructed CCDs
and CMDs. The goal is to get some information about the cluster
reddening, age and distance.
A hint of the cluster reddening derives from Schlegel et al. (1998) dust
emission reddening maps. In the direction of NGC~7036 the reddening is 0.08.
This is basically confirmed by the analysis of the $(B-V)$ vs $(U-B)$
diagram presented in Fig.~3.
Here we consider the stars brighter than $V$~=~17 (filled circles)
which lie very close to to the empirical Schmidt-Kaler(1982) ZAMS 
(solid line), and therefore are weakly 
reddened. The remaining stars (open squares) are much more dispersed.

   \begin{figure*}
   \centering
   \resizebox{\hsize}{!}{\includegraphics{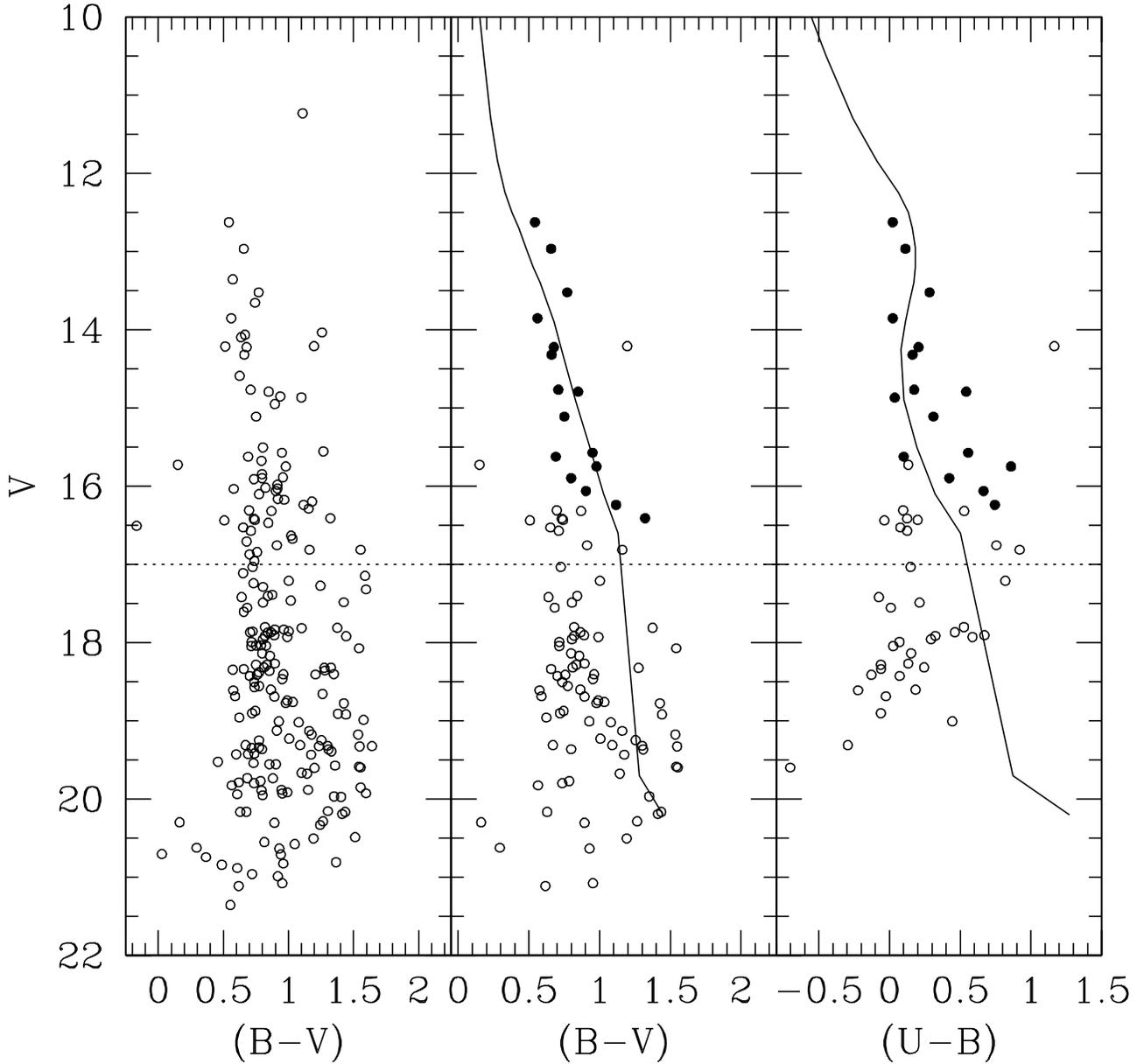}} 
   \caption{ CMDs of the stars in the region of NGC~7036. In the left
panel we plot all the stars. In the middle and right panel we consider
only the stars in the ring defined by $1.5 \leq r \leq 3.5$. The dotted
line provide a magnitude limit for probable cluster members. Finally,
the solid line is a ZAMS shifted by E$(B-V)$~=~0.1 and $(m-M)$~=~10.7.
Filled circles identify likely members.}
    \end{figure*}

\begin{table}
\tabcolsep 0.10cm
\caption{Photometry of likely member stars in the field of NGC~7036 deserving
further spectroscopic investigation.}
\begin{tabular}{cccccccc}
\hline
\hline
\multicolumn{1}{c}{ID} &
\multicolumn{1}{c}{$X(pixels)$}&
\multicolumn{1}{c}{$Y(pixels)$}&
\multicolumn{1}{c}{$V$}  &
\multicolumn{1}{c}{$(B-V)$} &
\multicolumn{1}{c}{$(U-B)$}  &
\multicolumn{1}{c}{$(V-I)$} \\
\hline
 2&     791.924&     535.611&     12.624&     0.543&    0.022&      0.662\\
 3&     705.509&     314.154&     12.965&     0.657&    0.112&      0.765\\
 5&     185.725&     494.985&     13.521&     0.771&    0.283&      0.878\\
 6&     105.387&     357.455&     14.866&    -0.440&    0.037&      2.009\\
 8&     256.450&     606.151&     13.854&     0.560&    0.023&      0.690\\
13&     265.737&     638.141&     14.221&     0.677&    0.204&      0.775\\
14&     102.390&     642.038&     14.318&     0.660&    0.162&      0.733\\
18&     200.052&     216.644&     14.767&     0.709&    0.175&      0.823\\
20&     230.678&     608.178&     14.791&     0.848&    0.542&      0.904\\
23&     273.431&     385.252&     15.109&     0.751&    0.310&      0.849\\
28&     876.273&     317.345&     15.571&     0.950&    0.556&      1.071\\
29&     624.969&     828.599&     15.622&     0.690&    0.101&      0.815\\
32&     876.329&     289.567&     15.747&     0.979&    0.858&      1.064\\
35&     588.382&     147.015&     15.897&     0.798&    0.422&      0.892\\
42&     716.759&     449.891&     16.061&     0.903&    0.665&      1.020\\
44&     413.532&     204.836&     16.239&     1.116&    0.744&      1.262\\
47&     569.217&     209.495&     16.410&     1.322&    1.641&      1.535\\
\hline
\end{tabular}
\end{table}

To guide the eye, we plotted also an empirical ZAMS (dashed line)
shifted by E$(B-V)$~=~0.1.\\
It seems that when looking for possible members one has to consider the
brightest stars. It is no possible on the other hand to select cluster members
by using individual reddenings, since the stars in Fig.~3 are probably
of spectral type later than $G$.\\
The same indication about the reddening derives from the analysis of the
$BVI$ photometry, following the method devised by Munari \& Carraro (1996),
which yields E$(B-V)$~=0.10$\pm$0.05.\\
In Fig.~5 we present three CMDs for the stars in the field of NGC~7036.
In the left panel we plot all the stars, in the middle panel
we plot the stars lying in the ring $1.5 \leq r \leq 3.5$
where we have seem that the brightest stars are confined. Finally, 
in the right panel we plot the same stars, but in the plane $(U-B)$ vs $V$
. The dotted line indicates a magnitude limit for probable members.
These latter are indicated with filled symbols.\\
It is very difficult to get an estimate of the cluster age and distance
since apparently there are no evolved stars. This can be explained 
statistically, since the cluster is intrinsically poorly populated, and 
therefore the absence of evolved stars is not completely unexpected.
The only star which stays in the evolved region of the CMD is also 
the brightest one, and is very probably a field star located between us and 
NGC~7036, since it appears projected apart from the cluster central
region.\\
To have a rough estimate of cluster distance we  have to rely only on MS stars,
and we proceed as follows.\\
From the location of stars in the $(B-V)$ vs $(U-B)$ plane, we infer
that the stars spectral types range from about $G0$ to $M2$ by deriving
the absolute colors from the ZAMS at the same position of the stars. 
This implies that the distance modulus is $(m-M) \approx 10.7\pm0.5$, which
corresponds to a distance of about 1 kpc.
Moreover, if the stars having $G0$ spectral type are still along
the Main Sequence (MS), we infer a probable age of about 3-4 Gyr.\\

In conclusion, we are tempted to suggest that NGC~7036 is indeed
an OCR having 17 likely members, whose properties are summarized in Table~3. 
Star counts seem to support this suggestion. CCDs and CMDs are
more difficult to interpret. Anyway, if NGC~7036 is a bound stellar
aggregate, what remains is an old, weakly reddened star cluster
1 kpc far from the Sun.

   \begin{figure}
   \centering
      \resizebox{\hsize}{!}{\includegraphics{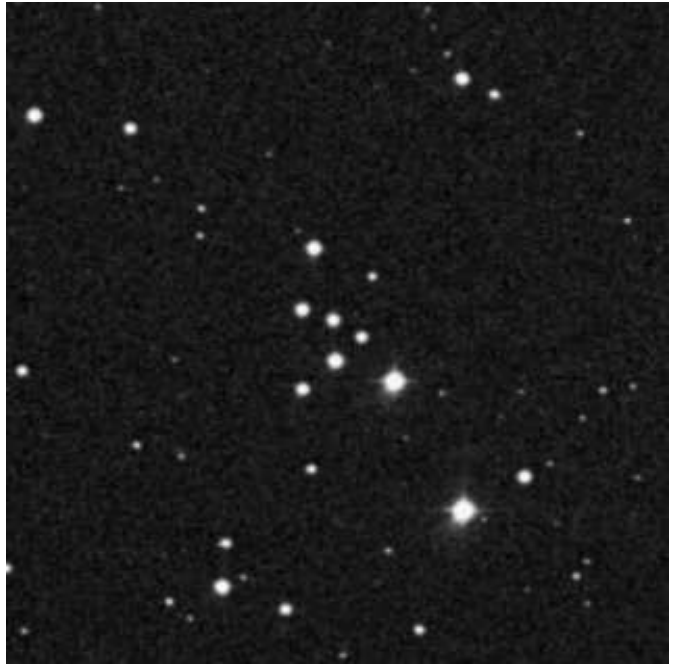}} 
   \caption{A red DSS map of the covered region in the field
of NGC~7772. North is up, East on the left.}
    \end{figure}
   \begin{figure}
   \centering
   \resizebox{\hsize}{!}{\includegraphics{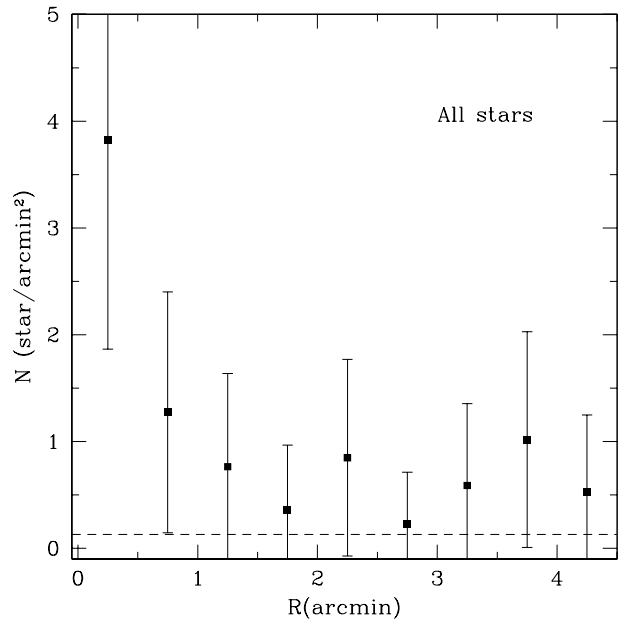}} 
   \caption{ Star counts in the field of 
of NGC~7772 as a function of the radius. The dashed line is the
field number density estimate provided by Bica et al. (2001).}
    \end{figure}

\section{NGC~7772}
NGC~7772 (OCL~230, see Fig.~6) appears as an asterism of 8 bright stars
in a poorly populated field. It resembles very 
closely NGC~6994 (Carraro 2000, Bassino et al. 2000). As for NGC~7036,
this cluster has never studied insofar.\\ 
In the field of the cluster
there is the star GSC~01722-01669 (TYC~1722~1669~1), 
$\#1$ in our numbering system,
which presumably has no relation with NGC~7772, since it lies outside
the cluster core. It is probably a blue star located between us and NGC~7772.

\subsection{Star counts}
We derived the surface stellar density by performing star counts
in concentric rings around stars \#8 (selected as 
approximate cluster center) and then dividing by their
respective surfaces. The final density profile and the corresponding
poissonian error bars are depicted in Fig.~7.
In this figure we take into account all the measured stars.\\
The surface density decreases sharply  up to a 
radius of about 2 arcmin, afterwards the density profile
it is basically flat. Therefore NGC~7772 appears to be very
compact with a core radius which at maximum amounts to 2 arcmin.\\
Also in this case star counts indicate that we are facing a  significant
density contrast with respect to the field, whose  
bright stars ($B$~=~13.5, $V$~=~13.0) density in this direction
is represented by the dashed line in Fig.~7. We count 5
stars brighter than $V~=13.0$ in a field of about 9 squared arcmin,
basically the same number reported by Bica et al. (2001, Table~1).

\subsection{Color-Color and Color-Magnitude Diagrams}
In order to better understand the nature of NGC~7772, we constructed CCDs
and CMDs. The goal is again to get information about the cluster
reddening, age and distance.
A hint of the cluster reddening derives from Schlegel et al. (1998) dust
emission reddening maps. In the direction of NGC~7772 the reddening is 0.04.
This is basically confirmed by the analysis of the $(B-V)$ vs $(U-B)$
diagram presented in Fig.~8, where all the stars having $UBV$ photometry 
are plotted. One can readily see that these stars lie along the empirical ZAMS
(solid line) taken from Schmidt-Kaler (1982).

\begin{table}
\tabcolsep 0.10cm
\caption{Photometry of likely member stars in the field of NGC~7772 deserving further
spectroscopic investigation.}
\begin{tabular}{cccccccc}
\hline
\hline
\multicolumn{1}{c}{ID} &
\multicolumn{1}{c}{$X(pixels)$}&
\multicolumn{1}{c}{$Y(pixels)$}&
\multicolumn{1}{c}{$V$}  &
\multicolumn{1}{c}{$(B-V)$} &
\multicolumn{1}{c}{$(U-B)$}  &
\multicolumn{1}{c}{$(V-I)$} \\
\hline
   2&   532.296&   603.489&   11.183&    1.310&    1.459&    1.399\\
   3&   636.041&   568.166&   12.529&    0.719&    0.262&    0.855\\
   4&   677.386&   371.724&   12.718&    0.578&    0.047&    0.741\\
   5&   831.308&   968.188&   12.662&    0.817&    0.554&    0.953\\
   6&   414.670&    75.715&   13.171&    0.553&   -0.071&    0.786\\
   7&   693.815&   479.542&   13.340&    0.929&    0.837&    1.031\\
   8&   640.231&   496.972&   13.461&    0.628&    0.102&    0.772\\
   9&   303.175&   769.777&   13.557&    0.880&    0.587&    1.018\\
  10&   691.480&   617.749&   13.635&    0.705&    0.286&    0.839\\
  11&   998.222&   163.777&   14.065&    0.715&    0.444&    0.831\\
  12&   590.952&   525.387&   14.242&    0.681&    0.199&    0.831\\   
  13&   719.036&  1003.786&   14.278&    0.743&    0.372&    0.864\\
  14&   827.791&   887.162&   15.209&    0.835&    0.543&    0.964\\
  15&   360.372&    99.613&   15.346&    0.532&   -0.077&    0.715\\
  16&   675.352&   758.966&   15.618&    0.997&    1.108&    1.072\\
\hline
\end{tabular}
\end{table}

   \begin{figure}
   \centering
   \resizebox{\hsize}{!}{\includegraphics{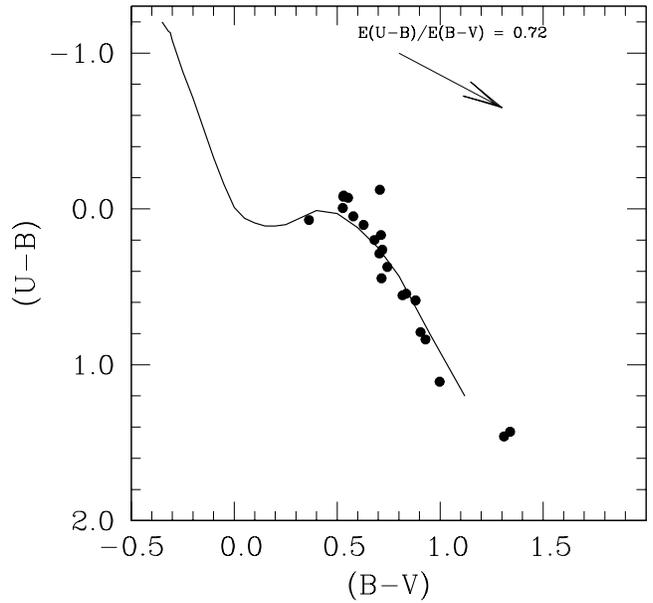}} 
   \caption{ Color-color diagram for all the stars in the field
of NGC~7772 having $UBV$ photometry. 
The solid line is a Schmidt-Kaler (1982) empirical ZAMS.}
    \end{figure}

   \begin{figure}
   \centering
   \resizebox{\hsize}{!}{\includegraphics{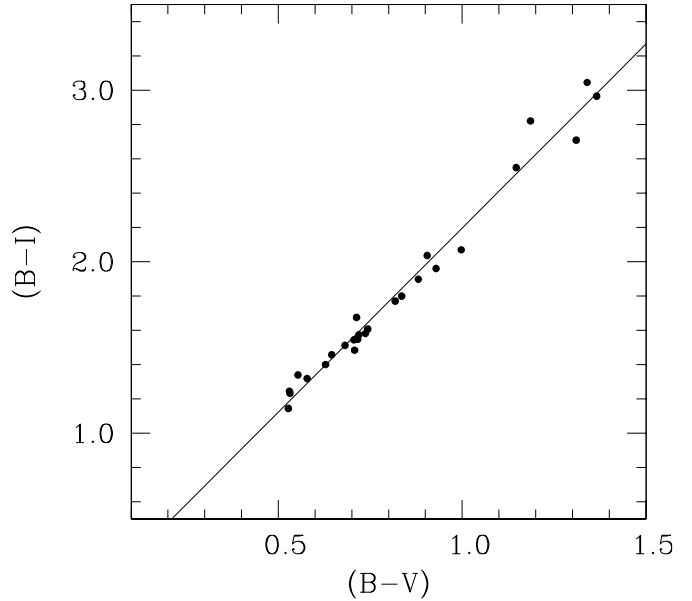}} 
   \caption{ NGC~7772 stars in the $(B-V)$ vs
$(B-I)$ plane.}
    \end{figure}
   
\noindent
The same result is obtained
by considering the distribution of the stars in the $(B-V)$ vs $(B-I)$
(see Fig.~9).
By applying the Munari \& Carraro (1996) method we infer a reddening
E$(B-V)$~=~0.05$\pm$0.03.\\
As in the case of NGC~7036,  we can infer the approximate spectral type by analysing
colours in the $(B-V)$ vs $(U-B)$ plane. It turns out that the spectral type
of possible members ranges from about $A8$ to $M3$.
In Fig.~10 we present three CMDs for the stars in the field of NGC~7772.
In the left panel we plot all the stars, in the middle panel
we plot the stars lying within $r \leq 2.5$
where we have seen that the brightest stars are confined. Finally, 
in the right panel we plot the same stars, but in the plane $(V-I)$ vs $V$. \\
Most of the stars in all the panels above $V$~=~16 are probable MS stars.
There is only an exception, which is star $\#2$, which lies within
the cluster core, and that we consider as a giant star probable 
member of the cluster.
In the three panels of Fig.~10 we have overlaid a solar metallicity
isochrone taken from Girardi et al. (2000) for the age of 1.5 Gyr.
This isochrone provides a good fit of the data down to $V$~=16.0 by
adopting E$(B-V)$~=~0.03 and $(m-M)$~=~11.1$\pm$0.3.
Below $V$~=~16.0, the MS starts to be ill defined. This is a clear 
signatures of low mass stars depletion, in nice analogy with -for instance-
the cases  of NGC~3680 (Anthony-Twarog et al. 1991) and NGC~7762
(Patat \& Carraro 1995).\\

In conclusion, we confirm previous suggestions
by Bica et al. (2001)  that NGC~7772 is an OCR 1.5 Gyr old. We identify
14 likely members, which are plotted with filled symbols in Fig.~10,
and whose properties are summarized in Table~4.

   \begin{figure*}
   \centering
   \resizebox{\hsize}{!}{\includegraphics{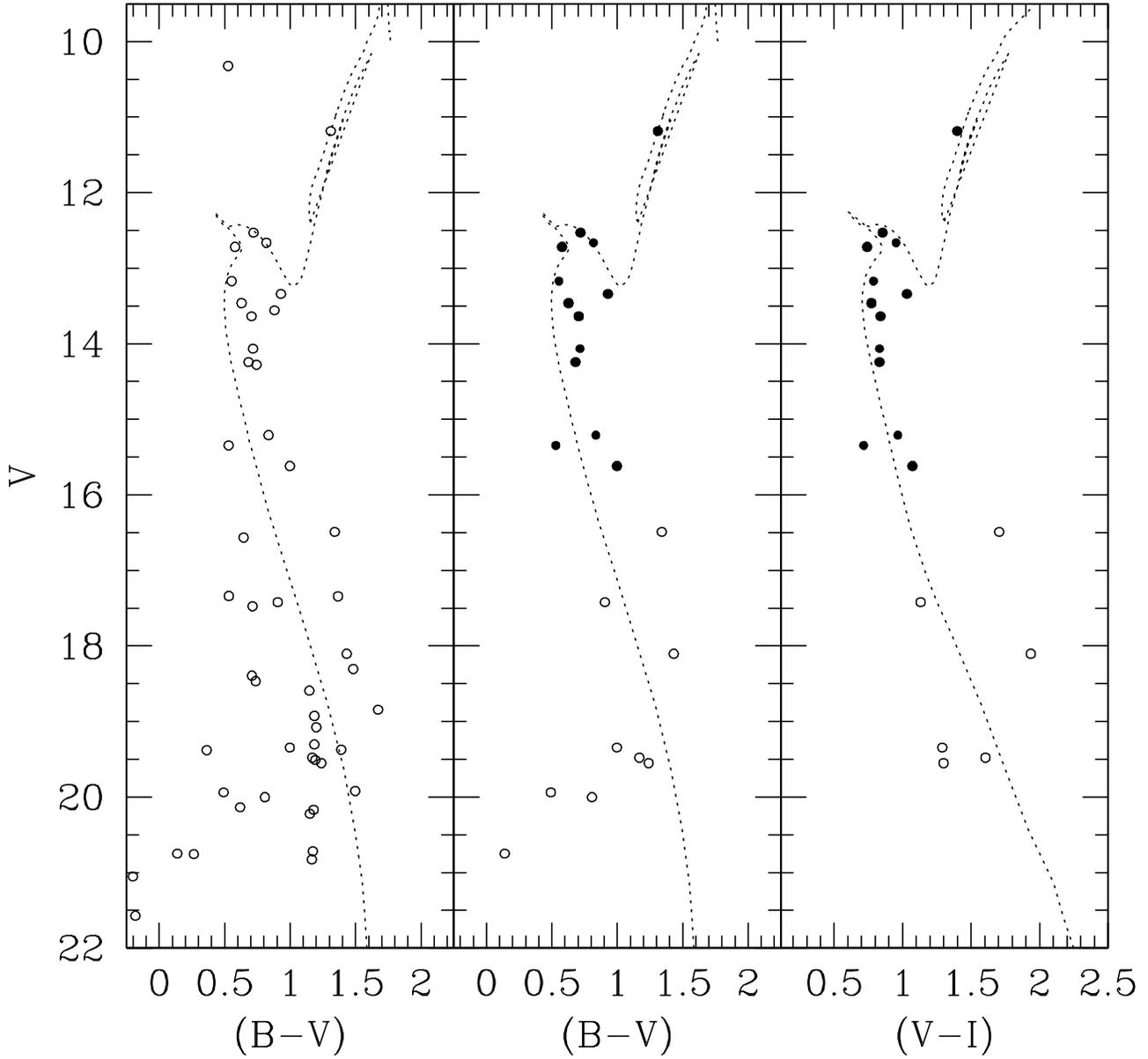}} 
   \caption{ CMDs of the stars in the region of NGC~7772. In the left
panel we plot all the stars in the $(B-V)$ vs $V$ plane. 
In the middle and right panel we consider
only the stars within $r \leq 2.5$. The dotted
line is a solar metallicity isochrone for the age of 1.5 Gyr
taken from Girardi et al. (2000), and
shifted by E$(B-V)$~=~0.03 and $(m-M)$~=~11.1.
Filled circles identify likely members.}
    \end{figure*}

\section{Conclusions}
We have presented the first CCD $UBVI$ observation of the Probable
Open Cluster Remnants NGC~7036 and NGC~7772.\\
Our analysis suggests that:

\begin{itemize}

\item NGC~7036 and NGC~7772 are two nice examples of intermediate-age
open clusters
in advanced stages of dynamical evolution;
\item NGC~7036 is a 3-4 Gyr open cluster located 1 kpc far from the Sun.
However, we stress the fact that this cluster still remains a 
somewhat doubtful  case, due to the absence of clear features in the CMDs;
\item NGC~7772 is a less doubtful case. We showed that the cluster underwent
strong low mass stars depletion. What remains is a group of 14 stars 1.5
Gyr old and located 1.5 kpc away from the Sun.
\end{itemize}

\noindent
It is worth to remarking that the present results must be considered 
with some caution, and that the list of members must be better
constrained by determining individual star radial velocities and proper
motions. This way these objects can become templates for N-body
simulation aimed at investigating the dynamical evolution of open
star clusters and the origin of the field star population

\begin{acknowledgements}
I am deeply indebted to Nando Patat and Sandro Villanova for numerous
very useful conversations. I acknowledge Giorgio Martonara
and Franco Bocchi for the kind night assistance. 
This study made use of Simbad. 
\end{acknowledgements}


\begin{thebibliography}{}

  \bibitem{} Anthony-Twarog, B.J., Heim, E.A., Twarog, B.A., \& Caldwell, 
             N., 1991, AJ 102, 1056
  \bibitem{} Bassino, L.P., Waldhausen, N. \& Mart\'inez, R.E. 2001, 
             A\&A 355, 138 
  \bibitem{} Baumgardt, H. 1988, A\&A 340, 402
  \bibitem{} Baumgardt, H., Dettbarn, C., Wielen, R. 2000,
             A\&AS 146, 251
  \bibitem{} Bergond, G., Leon , S., \& Guilbert, J., 2001, A\&A 377, 462
  \bibitem{} Bica, E., Santiago, B.X., Dutra, C.M., et al. 2001,
             A\&A 366, 827 
  \bibitem{} Carraro, G. 2000, A\&A 357, 145
  \bibitem{} Carraro, G., Patat, F. 1995, MNRAS 276, 563
  \bibitem{} De la Fuente Marcos, R. 1998, A\&A 333, L27
  \bibitem{} De la Fuente Marcos, R. 2001, in {\it Modes of Star
             Formation and the Origin of Field Population}, ASP 
             Conference Series, in press 
  \bibitem{} Desidera, S., Fantinel, D., Giro, E. 2001, AFOSC USER MANUAL
  \bibitem{} Dutra, C.M., Bica, E. 2000, A\&A 359, L9
  \bibitem{} Girardi, L., Bressan, A., Bertelli, G., Chiosi, C., 2001, A\&AS
              141, 371  
  \bibitem{} Landolt, A.U. 1992, AJ 104, 340
  \bibitem{} Lyng\aa, G., 1987, Catalog of Open Star Cluster Data, 
             Strasbourg, CDS
  \bibitem{} Munari, U., Carraro, G., 1996, A\&A 314, 108     
  \bibitem{} Odenkirchen, M., Sourinac, C.  2002, A\&A in press
  \bibitem{} Patat, F., Carraro, G. 1995, A\&AS 114, 281 
  \bibitem{} Patat, F., Carraro, G. 2001, MNRAS 325, 1591
  \bibitem{} Pavani, D.B., Bica, E., Dutra, C.M.,  et al. 2001, 
             A\&A 374, 554
  \bibitem{} Piatti, A.E., Clari\'a, J.J. 2001, A\&A 379, 453
  \bibitem{} Schlegel, D.J., Finkbeiner, D.P., \& Davis, M., 1998, ApJ 500, 
             525
  \bibitem{} Schmidt-Kaler, Th., 1982, Landolt-B\"ornstein,
             Numerical data and Functional Relationships in Science and
             Technology, New Series, Group VI, Vol. 2(b), K. Schaifers
             and H.H. Voigt Eds., Springer Verlag, Berlin, p.14   
\end{thebibliography}
\end{document}